Controlled interconversion of quantized spin wave modes via local magnetic fields.

Zhizhi Zhang[1,2], Michael Vogel[1], José Holanda[1], Junjia Ding[1],
M. Benjamin Jungfleisch[3], Yi Li[1,4], John E. Pearson[1], Ralu Divan[5], Wei Zhang[4],
Axel Hoffmann[1], Yan Nie[2,*], and Valentine Novosad[1,*]

[1] Materials Science Division, Argonne National Laboratory, Argonne, IL 60439, USA
[2] School of Optical and Electronic Information, Huazhong University of Science and Technology, Wuhan 430074, China
[3] Department of Physics and Astronomy, University of Delaware, Newark, DE 19716, USA
[4] Department of Physics, Oakland University, Rochester, MI 48309, USA
[5] Center for Nanoscale Materials, Argonne National Laboratory, Argonne, IL 60439, USA

Abstract

In the emerging field of magnonics, spin waves are considered for information processing and transmission at high frequencies. Towards this end, the manipulation of propagating spin waves in nanostructured waveguides for novel functionality has recently been attracting increasing focus of research. Excitations with uniform magnetic fields in such waveguides favors symmetric spin wave modes with odd quantization numbers. Interference between multiple odd spin wave modes leads to a periodic self-focusing effect of the propagating spin waves. Here we demonstrate, how antisymmetric spin wave modes with even quantization numbers can be induced by local magnetic fields in a well-controlled fashion. The resulting interference patterns are discussed within an analytical model and experimentally demonstrated using microfocused Brillouin light scattering (μ-BLS).

Introduction

Collective excitations of the electronic spin structure known as spin waves and their quasiparticles, i.e. magnons, are promising for high frequency information processing and transmission.[1-4] Additional functionality can be gained from the fact that spin waves can also be coupled to other wave-like excitations, such as photons[5, 6] and phonons.[7] Furthermore, many classical wave phenomena, such as diffraction,[8, 9] reflection and refraction,[10-12] interference[13, 14] and the Doppler effect[15, 16] were observed with spin waves. At the same time, quantum mechanical interactions, such as the magnon scattering[17-19] and their interactions with other quasiparticles[20] were observed as well, and provide additional avenues for utilizing spin waves. Understanding these phenomena is key to realizing practical applications in the rapidly emerging field of magnonics.

Spin waves can encode information either in their amplitude[21, 22] or their phase.[23, 24] Compared with conventional electronic approaches, spin waves possess several advantages, including potentially reduced heat dissipation,[25] wave-based computation[26,

[27] and strong nonlinearities,[28, 29] which may all be beneficial for efficient data processing. The recent emerging interest in magnonics can be attributed to the improvement of modern micro-fabrication, which enables the realization of the magnetic microstripes with characteristic dimensions ranging from several μm to below hundred nm[30, 31], as well as integrated micro-antenna for excitations[32, 33]. When such a magnetic microstripe is magnetized with an external magnetic field ($H_{ext}$) in-plane and perpendicular to the stripe direction, the spin waves are called Damon-Eshbach modes[34] and can be localized either at the edge or in the center region, depending on their frequencies[35, 36]. Previous studies demonstrated that spin waves at the center region (so-called waveguide spin waves), are quantized into several discrete modes due to the confinement along the width of the waveguide.[37] In addition, generally a homogenous rf field can only excite lateral symmetrically-distributed, odd waveguide spin wave modes.[38] The interference of several of these modes results in a periodic self-focusing, where the waveguide spin waves propagate in diamond chain-like channels[32, 39, 40].

In magnonic applications, the manipulation of the spin wave propagation is of great significance for the functionality of such devices, especially for logic elements[21-24] and multiplexers[41]. Towards this end constructive or destructive interference of multiple, coherent spin waves impact the spatial intensity distributions of the resultant waves, and therefore controls the energy and information flows associated with the spin waves. Previous investigations focused mostly on odd spin wave modes, since they were easier to generate with homogeneous excitations. In this work, we demonstrate the controlled interconversions of odd and even waveguide spin waves in yttrium iron garnet (YIG) microstripes by breaking the symmetry via well-defined local inhomogeneous magnetic fields. This allows for a reconfigurable mechanism of mode conversion, unlike previous experiments where the symmetry is broken by the geometry of the waveguide.[42] The local magnetic fields are generated from permalloy (Py, $Ni_{81}Fe_{19}$) micro-magnets placed asymmetrically next to the YIG waveguide. Note that the saturation magnetization ($M_s$) for permalloy is about five times larger than that for YIG. Using a combination of theoretical calculations, magnetic simulations, and microfocused Brillouin light scattering (μ-BLS), we demonstrate that the different spin wave channels are essentially controlled by the phase difference between odd and even modes, which can be practically modulated through the relative position of the micro-magnets and the magnitude of the external magnetic field.

Analytical Calculations

We consider a thin YIG microstripe with the thickness $t$ = 50 nm, width $w$ = 3 μm and infinite length $l$, magnetized in-plane in a direction perpendicular to the length through a magnetic field $H_0$ = 650 Oe as shown in the inset of Fig. 1 (a). The material parameters used in the theoretical calculation are $M_s$(YIG) = 1960 G, exchange constant $A$(YIG) = 4×10$^{-7}$ erg/cm, and damping factor $\alpha$(YIG) = 7.561×10$^{-4}$.[31]

For the first step, the waveguide spin wave modes in a microstripe can be described based on the dipole-exchange theory of the spin wave dispersion spectra in a continuous magnetic film.[43, 44] This theory provides explicit relations between the wave vector

$\boldsymbol{k} = (k_x, k_y)$ and the frequency $f$ of the spin waves:

$$f = \gamma \sqrt{\left(H_0 + M_s\left(1 - p + \lambda_{ex}^2 k^2\right)\right) \times \left(H_0 + M_s\left(p\frac{k_x^2}{k^2} + \lambda_{ex}^2 k^2\right)\right)} \quad , \tag{1}$$

where $p = 1-(1-e^{-kt})/kt$, $k^2 = k_x^2 + k_y^2$, and $\lambda_{ex} = (2A/M_s^2)^{1/2}$ is the exchange length.[45] The two limiting relations for $k_x = 0$ and $k_y = 0$ correspond to Demon-Eshbach and backward volume modes. Furthermore, there are scientific constants for the gyromagnetic ratio $\gamma = 2.8$ MHz/Oe.

Neglecting the effect of the demagnetizing field ($H_d$), which is important only close to the edges of the microstripe, the waveguide spin waves are confined along the width direction and can be described as the quantization of planar spin waves propagating along the length direction. It means that only the waveguide spin waves with $k_y$ components satisfying the resonant standing waves conditions can propagate in the microstripe. These $k_y$ components are a set of discrete values, described by a simple expression:

$$k_{y,n} = n\pi/w \quad . \tag{2}$$

Combining Eqs. (1) and (2), the dispersion relation curves for each mode with $n = 1,2,\ldots,5$ are plotted in Fig. 1(a). Only lateral modes with odd quantization numbers $n$ can be excited under uniform rf magnetic field, and their amplitudes decrease with increasing $n$ as $1/n$.[38] With a frequency of $f = 4$ GHz we can calculate the corresponding $k_{x,n}$. Then, the spatial distribution of the $n$th mode's dynamic magnetization and their integrated superpositions, i.e. the interference of the odd modes can be written as

$$m_n(x, y) \propto \frac{1}{n}\sin\left(\frac{n\pi}{w} y\right)\cos\left(k_{x,n} x - 2\pi f t + \varphi_n\right) \quad , \text{and} \tag{3}$$

$$I_\Sigma(x, y) \propto \left(\sum_n m_n(x, y)\right)^2 \quad , \tag{4}$$

where $\varphi_n$ is the excitation phase. The patterns of the first three odd modes are mapped in Fig. 1(b) for $-2\pi ft + \varphi_n = 0$, which coincides with the maximum dynamic magnetization at $x = 0$. According to Eqs. (3) and (4), the major contribution to $I_\Sigma(x,y)$ comes from the first few modes, since the intensity of the modes are proportional to $1/n^2$. Therefore, $n = 11$ is sufficient for an accurate analysis and the corresponding interference pattern is mapped as shown in the upper panel of Fig. 1(c). In order to determine the amplitude of the procession of every spin, we calculated the maximum values of $I_\Sigma(x,y)$ within $-2\pi ft + \varphi_n \in (0, 2\pi)$:

$$I(x, y) = \max\left[I_\Sigma(x, y) : -2\pi ft + \varphi_n \in (0, 2\pi)\right] \quad , \tag{5}$$

where $I(x,y)$ is the amplitude of the waveguide spin wave in materials (without considering damping effects), which can be detected using the μ-BLS technique. The waveguide spin wave intensity pattern for odd numbers $n$ is mapped in the lower panel of Fig. 1(c). It shows that the interference of the odd modes results in a symmetric rhombohedral-shaped channel. Here, it should be pointed out that mathematically the phase differences of the lower modes ($n=1, 3$) between the adjacent nodes (I, II and III in Fig. 1(c)) of the spin wave pattern are approximately $2q\pi+\pi$, where $q$ is an arbitrary integer, as shown in Fig. 1(d).

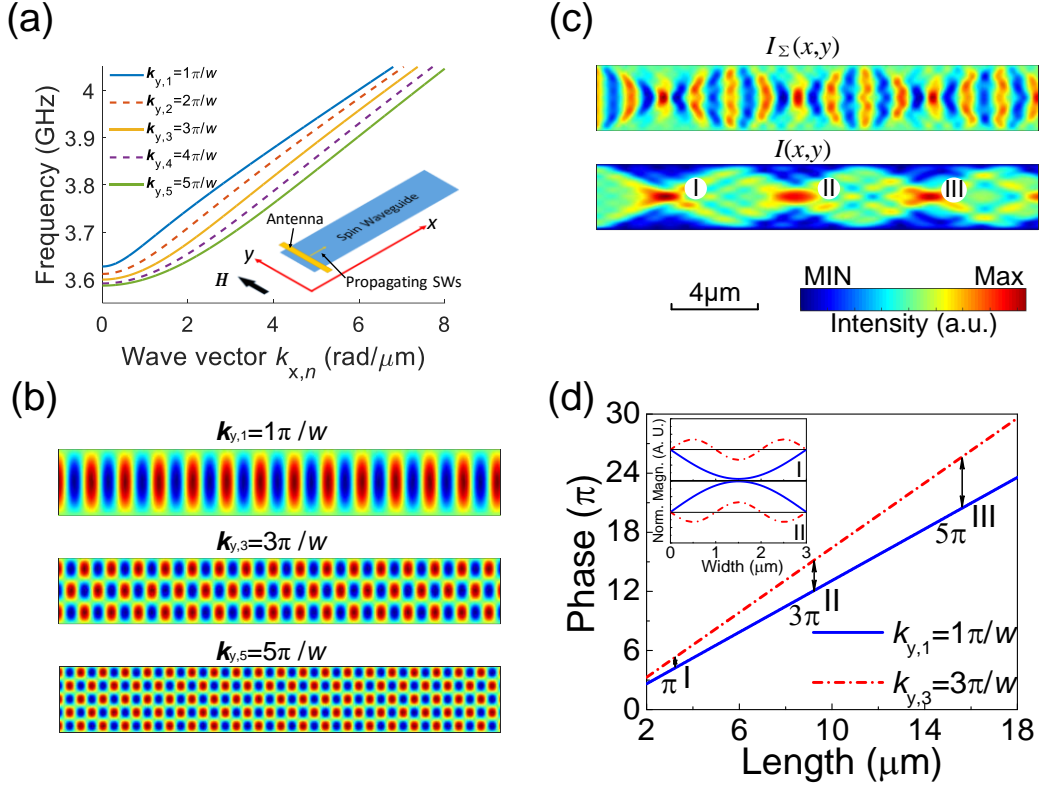

Fig. 1 Theoretical calculated results: (a) Dispersion curves for the first five Damon Eshbach mode waveguide spin waves propagating in a YIG microstripe. The inset depicts a schematic of the studied model. (b) The spatial distribution of $m_n(x,y)$ for the first three odd modes at the initial phase ($-2\pi ft+\varphi_n = 0$). (c) Interference patterns of the first few odd modes ($n \le 11$), upper panel: $I_\Sigma(x,y)$ and lower panel: $I(x,y)$. (d) The phase shift of the first odd modes along the length. Inset shows the normalized dynamic magnetization distribution across the stripe at the first and second nodes as indicated in (c).

Introducing new modes to interfere with the existing modes should modify this flow pattern. Towards this end, we consider the even modes because: 1. they have the same frequency as the previously considered odd modes, and therefore the coherent interference would lead to a time-invariant pattern; 2. they should be easy to excite and should have comparable lifetimes compared to the odd modes in the waveguides. In contrast to the odd modes, the even modes have antisymmetric patterns; in other words, $m_n(x,y) + m_n(x,w-y) = 0$ for even $n$ according to Eq. (3). The patterns of the first two even modes are mapped in Fig. 2(a).

The interference patterns are strongly depended on the difference of the initial phases ($\Delta\varphi=\varphi_{odd}-\varphi_{even}$), which means that the waveguide spin wave channels can be controlled through tuning $\Delta\varphi$ between the odd and the even modes. For our analysis, some representative values (0, $\pi/2$, $\pi$, and $3\pi/2$) for $\Delta\varphi$ were chosen by fixing $\varphi_{odd} = 0$ in Eq. (3), and changing $\varphi_{even} = 0$, $\pi/2$, $\pi$, and $3\pi/2$, respectively. The corresponding patterns of $I_\Sigma(x,y)$ and $I(x,y)$ are shown in Fig. 2(b)-(e). Compared with Fig. 1(c), the introduction of the new modes changes the patterns from symmetric diamond-like shapes to antisymmetric zig-zag shapes. In addition, the paths of the waveguide spin waves can be continuously changed if $\Delta\varphi$ is varied continuously in the range from 0 to $2\pi$. Since the phase shift is given by $\Delta\varphi = kd$, we investigated the control of the $\Delta\varphi$ via two different pathways: the change of distance $d$, and the wave vector $k$.

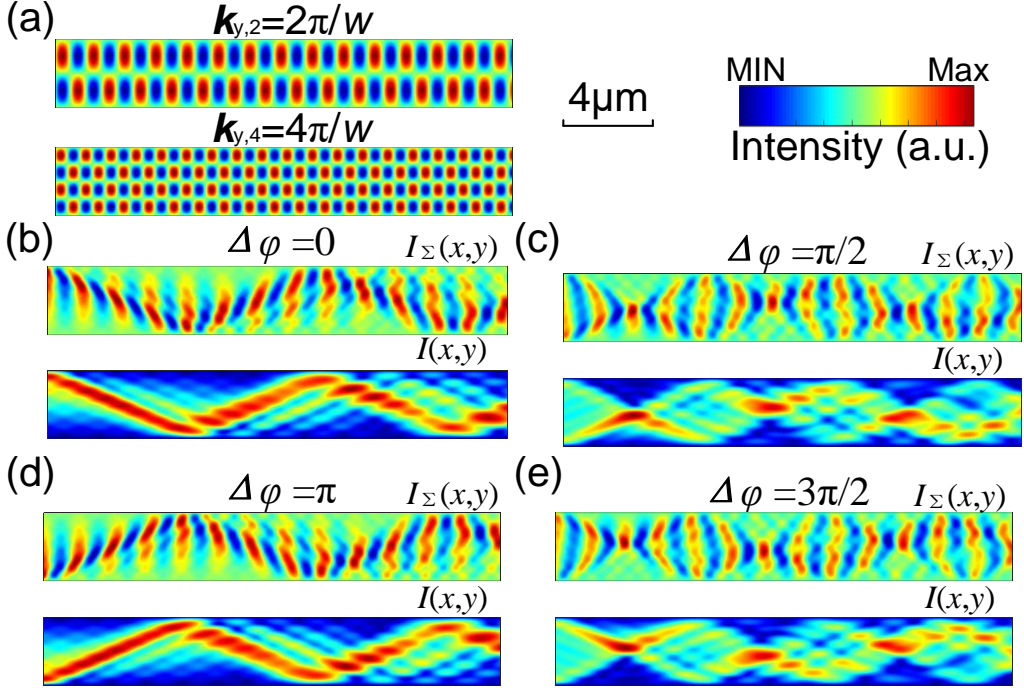

Fig. 2 (a) The spatial distribution of $m_n(x,y)$ for the first two even modes at the initial phase ($-2\pi ft+\varphi_n = 0$). Interference patterns of the odd and even modes with phase difference (b) $\Delta\varphi = 0$, (c) $\pi/2$, (d) $\pi$, and (e) $3\pi/2$, upper panel: $I_\Sigma(x,y)$ and lower panel: $I(x,y)$.

Effect of the distance $d$

In the discussions above, the introduction of even modes allows to manipulate the propagating waveguide spin waves through their interference with the intrinsic odd modes. The generation of even modes can be realized via the breaking of translational symmetry, for example, by passing through curved waveguides[42, 46]. In this work, we demonstrate that the magnetic symmetry of the single YIG microstripe can be broken by non-symmetric distribution of lateral micro-magnets, i.e., a permalloy dot as shown in Fig. 3(a). The simulations were performed using MuMax3[47]. The material parameters for permalloy (Py) are $M_s$(Py) = $1.08\times10^4$ G, $A$(Py) = $1.3\times10^{-11}$ J/m and $\alpha$(Py) = 0.01.[48] The external magnetic field ($H_{ext}$) set in the simulation was 640 Oe. The y component of the static effective magnetic field ($H_{eff}$) distribution inside of the YIG microstripe is shown in the color map of Fig. 3(a). Due to the strong induced dipolar field, the lateral symmetry of $H_{eff}$ across the width of the waveguide is gradually broken in the segment close to the permalloy dot, while $H_{eff}$ is again symmetric in the segments far away from the permalloy dot. For exciting the spin waves, we apply a continuous excitation of "sin" function $h_x = h_0\sin(2\pi f t)$ in the antenna region, with $f = 4$ GHz, and $h_0 = 1$ Oe, which is weak enough to avoid nonlinear effects. The total simulation time was 80 ns, to ensure that the system reaches a steady state. Fig. 3(b) shows the pattern of the waveguide spin waves in a single YIG microstripe, which is similar to the theoretical result in Fig. 1(c). Note, that the length of the spin wave modulation period in the simulation is slightly different to the ones previously calculated analytically, which is due to the reduced effective width by the demagnetic field and the slightly different $H_{ext}$.

Fig. 3(c) to (f) show the propagating waveguide spin wave patterns when the permalloy dot was located at the first node, antinode, the second node, and antinode respectively. They are qualitatively in accordance with the patterns of $\Delta\varphi = \pi$, $3\pi/2$, 0 and $\pi/2$ in Fig. 2. Practically, the odd modes are excited in the antenna region, with $\varphi_{odd} = 0$. As the odd modes propagate along the stripe for a certain distance $d$, the phases shift by $kd$, where $k$ is the corresponding wavevectors. At the first node position, the phase shift of the main contributing odd modes is approximately $\varphi_{odd} = 2q\pi+\pi$ as discussed above. Here, since the symmetry is broken, the even modes are excited with $\varphi_{even} = 0$, and therefore, the final interference pattern in Fig. 3(c) agrees well with the analytical result of $\Delta\varphi = \pi$. Similarly, the patterns of Fig. 3(d) to (f) agree with $\Delta\varphi = 3\pi/2$, 0 and $\pi/2$, respectively.

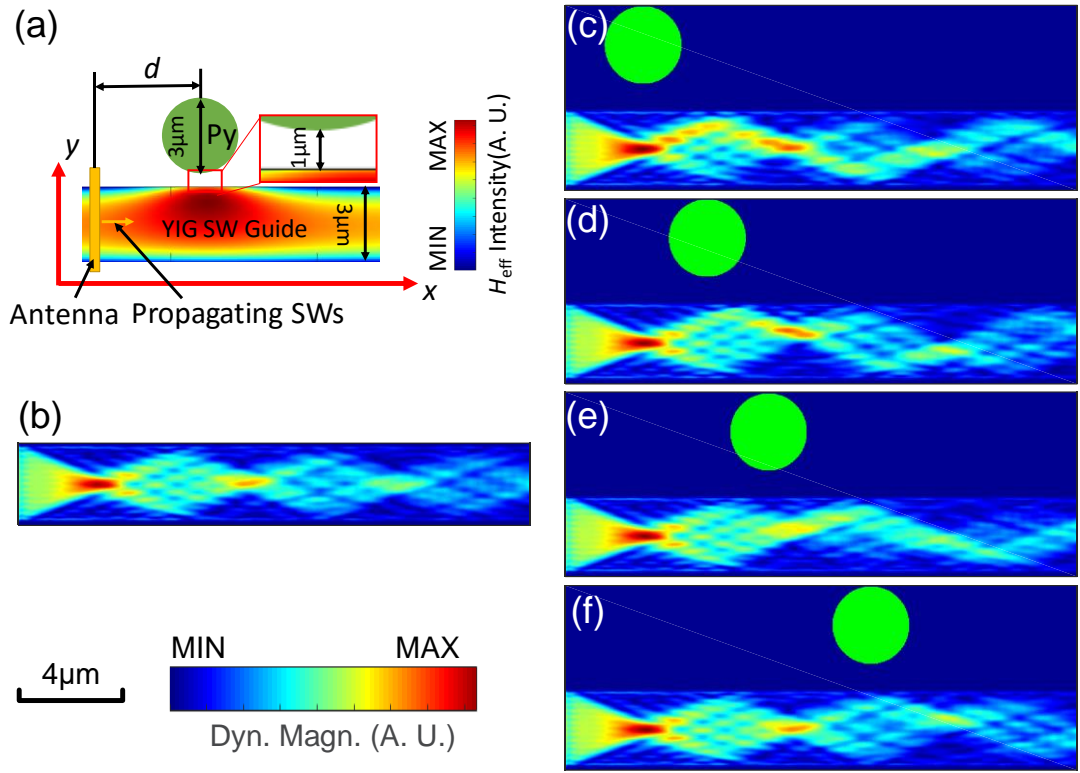

Fig. 3 (a) Schematic of the simulated model. (b) The *y* component of the effective magnetic field ($H_{eff}$) distribution inside of the YIG stripe with a permalloy (Py) dot (green, same hereinafter). Patterns of the waveguide spin waves propagating in (c) single YIG stripe, and YIG stripe with a lateral permalloy dot at the (d) first node, (e) first antinode, (f) second node and (g) second antinode.

In addition, it should be pointed out that the initial phase of the newly introduced even modes is also determined by which side the permalloy dot is located on. For example, comparing Fig. 3(c) and (e), the patterns of the waveguide spin waves after passing by the permalloy dot are inversely mirrored. Similar behavior is also observed for Fig. 3(d) and (f). This indicates that a phase difference of π can be induced by placing the permalloy dot on the other side. Therefore, the even modes can be annihilated (enhanced) by the destructive(constructive) interference with other even modes generated by other micromagnet in close proximity to the waveguide on the same (other) side one period away. In order to demonstrate this, we simulated the waveguide spin wave patterns in a YIG microstripe with three permalloy dots distributed on one side and two sides as shown in Fig. 4(a) and (b). In Fig. 4 (a), the permalloy dots were located at the first three nodes on one side. The waveguide spin waves experienced the following processes: 1. the first even mode (EM1) was generated with $\varphi_{EM1} = 0$ at the first node, resulting in the waveguide spin waves propagating non-symmetrically in the following self-focusing period; 2. the second even mode (EM2) was generated with $\varphi_{EM2} = 0$ at the second node. However, at this point, the first even mode has a phase shift of $\pi$ and destructively interferes with the second even mode. Therefore, the asymmetry disappeared in the next period; 3. the third even mode (EM3) was generated with $\varphi_{EM3} = 0$ at the third node again, leading to the following asymmetrical pattern. On the contrary, in Fig. 4(b), the second even mode was generated with $\varphi_{EM2} = \pi$ and thus constructively interfered with the first even mode, as did the third even mode. The anti-symmetric component was therefore increased compared with Fig. 3(c).

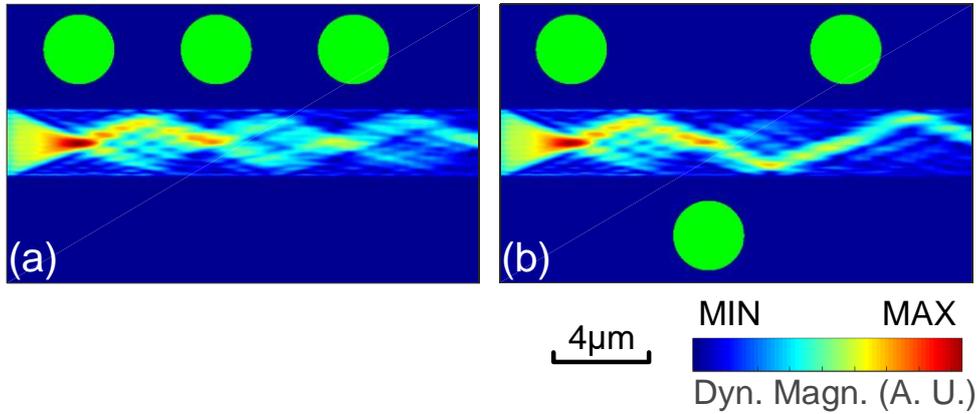

Fig. 4 Simulated patterns of the waveguide spin waves propagating in YIG stripe with three lateral permalloy dots at the first three nodes (a) on one side and (b) with the second permalloy dot on the opposite side.

In this section we demonstrated that $\Delta\varphi$ can be tuned by changing the relative position of permalloy dots near the YIG microstripe, including the distance $d$ to the excitation, and the side on which it is located. Changing the distance $d$ leads to a phase shift of odd modes with $kd$, and switching the sides cause even modes phases to shift by $\pi$. Using multiple permalloy dots introduces multiple even modes, whose constructive (destructive) interference increases (decreases) the anti-symmetric component of the propagating waveguide spin waves.

Effects of the wave vector $k$

According to the dispersion relation described by Eq. (1), the wave vectors $k$ of the waveguide spin waves with specific frequencies can be modified by $H_0$, which is the most common tunable parameter among the variables in the equation if the devices are already fabricated.[49, 50] Fig. 5(a) shows a schematic illustration of the investigated device, which is a 4.5-μm wide YIG (75-nm thick) stripe. The fabrication of the structures was done using electron-beam lithography and lift-off. For the excitation of the spin waves, the shortened end of a coplanar waveguide made of Ti(20 nm)/Au(500 nm) with a width ~2μm was placed on top of the end of the YIG microstripe. The spin waves excited by the antenna structure connected with a microwave generator can reach in several GHz frequency range. In this work, we fixed the frequency at 4 GHz. All the observations of the spin waves were performed using microfocused Brillouin light scattering (μ-BLS)[51] with a laser wavelength of 532 nm. First, we measured the 4 GHz spin wave intensity versus $H_{ext}$ in a single YIG stripe with the laser spot fixed at the center of the cross in the red circle as indicated in Fig. 5(a). The BLS intensities versus magnetic field is shown as in Fig. 5(b), where the peak is located around 650 Oe. It means that the 4-GHz spin waves propagate with the highest efficiency in the YIG microstripe for $H_{ext} \approx 650$ Oe. Subsequently, the intensity patterns of propagating spin waves in a single YIG microstripe under 630 and 670 Oe were mapped as shown in Fig. 5(c) and (d). Comparing the two patterns in a single YIG microstripe, the self-focus period was expanded with the increase of $H_{ext}$ due to the collective decrease and the convergence of the $k$s for odd modes.[52, 53]

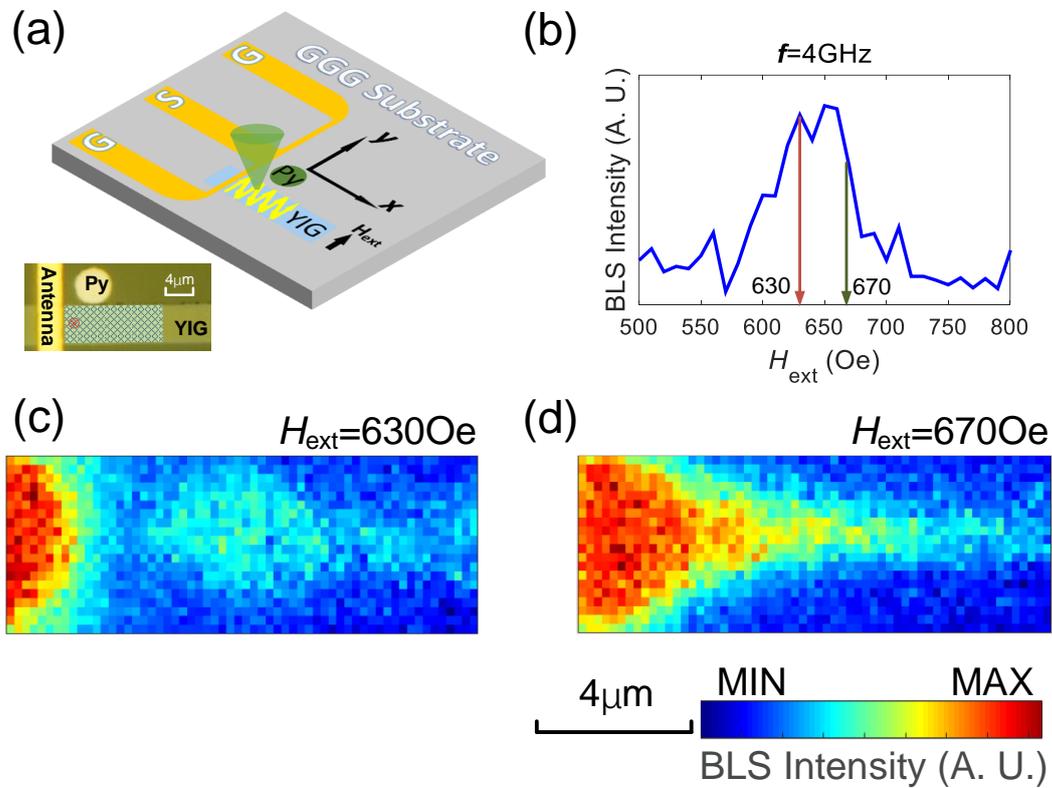

Fig. 5 (a) Schematic illustration of the device layout. The inset shows an optical microscopy image of the device. The spin wave patterns were imaged in the grid region. (b) 4-GHz spin wave BLS intensity in a single YIG stripe vs. $H_{ext}$ measured with the laser spot fixed at the center of the cross in the red circle. (c) and (d) BLS intensity images at two different applied fields.

Subsequently, the 4.5-μm permalloy dot was deposited using a combination of e-beam lithography and sputter deposition (see supplementary for experiment details), laterally on one side of the YIG microstripe ~3.5-μm away from the antenna, almost at the first node of the pattern measured for 630 Oe. Lastly, the spin wave intensities were imaged in the same region of the YIG microstripe under various magnetic field (610 to 690 Oe) as shown in Fig. 6.

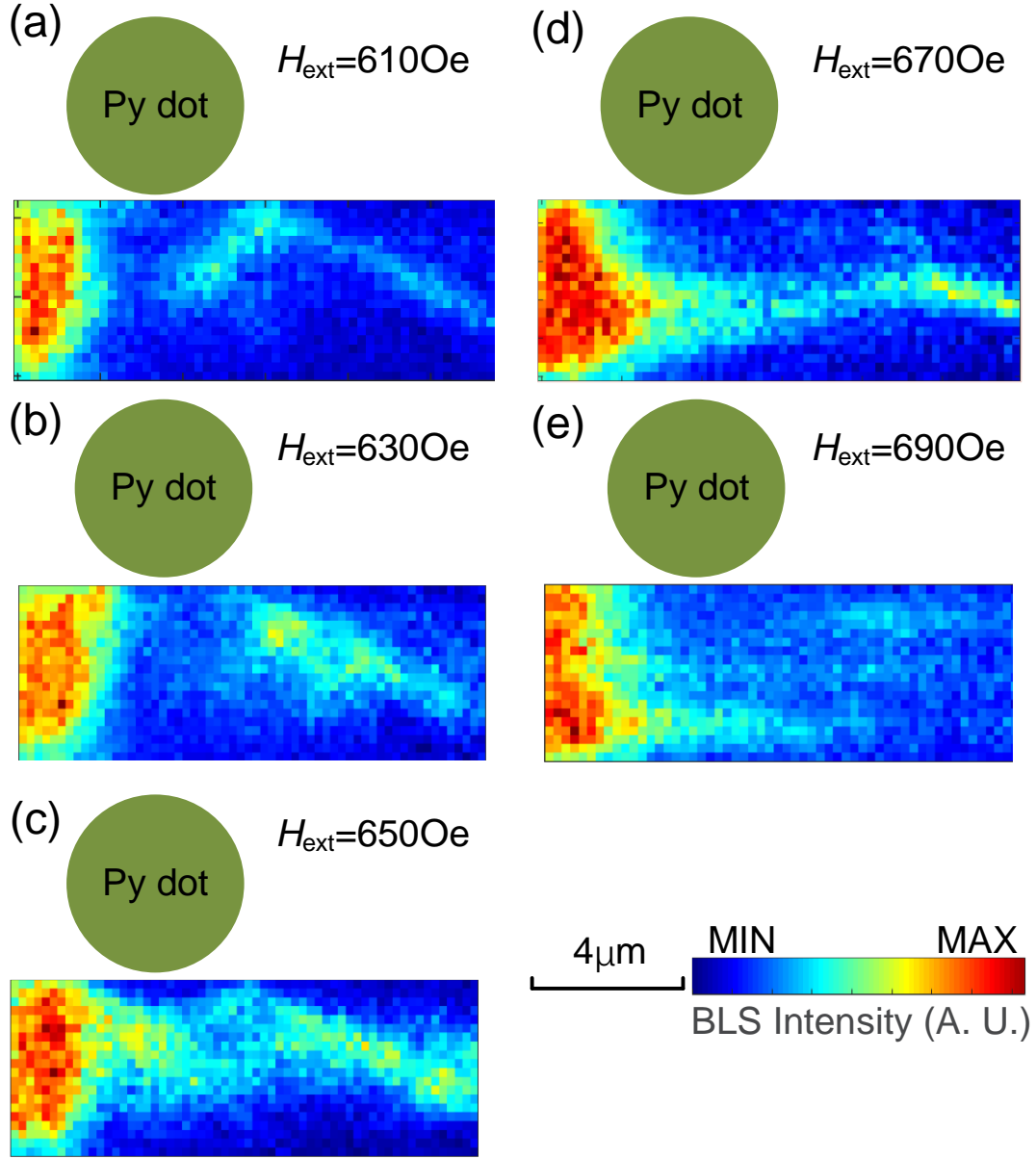

Fig. 6 4-GHz spin wave intensity patterns in YIG microstripe with a lateral permalloy dot measured at externally applied magnetic fields of (a) 610, (b) 630, (c) 650, (d) 670 and (e) 690 Oe.

  The BLS patterns in the YIG stripe without/with permalloy dot under 630 Oe [Fig. 5(c) and Fig. 6(b)] are in accordance with Fig. 3(b) and (c), where the spin waves flows toward the permalloy dot. On the contrary, comparing the patterns of Fig. 6(b) and (d), the effect of the permalloy dot at 670 Oe is to squeeze the spin wave flow toward the other side instead of attracting to the same edge, which indicates that the generated even modes here have a π phase difference with those in Fig. 6(b). According to Fig. 5(b), the 4-GHz spin waves propagate with the largest amplitude in the middle of YIG microstripe under $H_{ext}$ ≈650 Oe. The spin waves with a specific frequency in the waveguide could reach the highest intensity near the ferromagnetic resonant field. Similar phenomena were observed in measurements of the spin waves localized at the two edges of a stripe. The two SWs beams were split more with the increase of the field at a fixed frequency,[54] as well as the decrease of the frequency at a fixed field[35] due to the demagnetizing magnetic field. In order to demonstrate this effect, the $H_{eff}$ across the YIG stripe versus its width are plotted in Fig. 7(a), where the black dash line indicates the level of 650 Oe. The integrated BLS normalized intensities across the width close to permalloy dot were measured for different magnetic fields as shown in Fig. 7(b). The intersections between the dash line and solid lines in Fig. 7(a) agree with the locations of the BLS intensity peaks in Fig. 7(b) for the different magnetic fields. The presence of the permalloy dot introduces an additional static dipolar field, which shifts the position of the effective field being 650 Oe closer to (further away from) the permalloy dot when $H_{ext} < 650$ Oe ($H_{ext} > 650$ Oe), attracting (repelling) the spin wave flow.

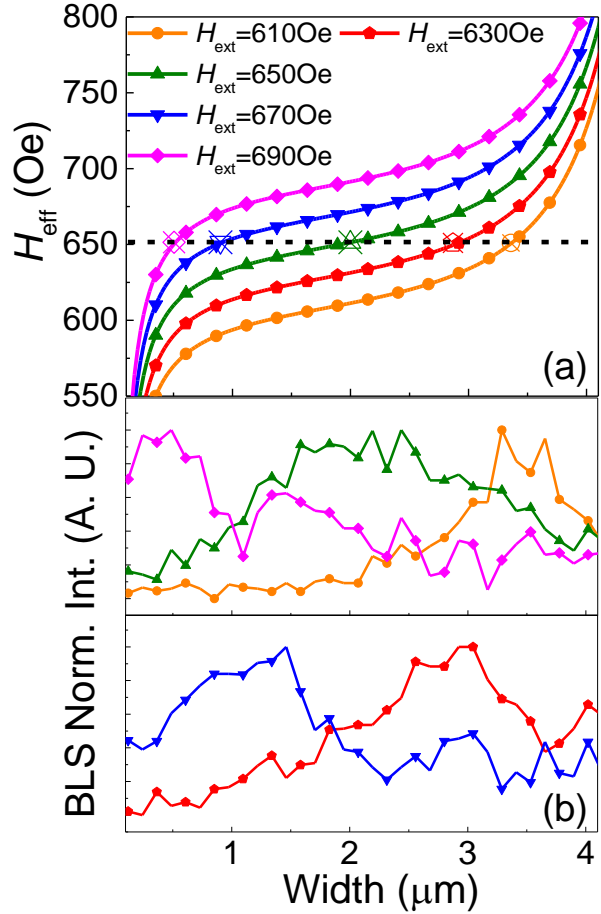

Fig. 7 (a) Simulated $H_{eff}$ and (b) integrated BLS normalized intensities across the YIG microstripe with a permalloy dot nearby at different $H_{ext}$ varied from 610 to 690 Oe. Horizontal black dash line in (a) indicates the field of 650 Oe. The intersections between the black dash line and the solid lines agree with the BLS intensity peaks, respectively. (a) and (b) share the same legend.

Conclusion

In summary, we demonstrated a new method, using interference of different spin waves, to manipulate the channels of the waveguide spin waves propagating in a magnetic microstripe. The waveguide spin wave channels can be tuned by the phase difference $\Delta\varphi$ between the intrinsic odd modes, which are preferred by homogenous excitation. Additional even modes can be introduced via breaking the magnetic symmetry through the non-symmetrical placement of a permalloy dot next to the wave guide. The phase shift $\Delta\varphi$ is controlled by the relative position of the permalloy dot to the antenna and the external magnetic field $H_{ext}$. An additional phase difference of $\pi$ can be introduced if the permalloy dot is located on the opposite side of the microstripe or the $H_{ext}$ exceeds the field for the most efficient spin wave propagation. These findings will assist with magnonic engineering, such as the design of a multiplexer combined with piezoelectric strain control of the micro-magnets. They might also enable new functionality, such as the non-reciprocity. Furthermore, note that with the suitable design of additional magnetic structures with sufficiently high anisotropy, the additional stray field may be modulated in a bistable manner, which could provide additional possibilities for controlling spin wave propagation. Lastly, this model system also serves as an ideal system for fundamental scientific research on the physics of wave propagation.

*Acknowledgment*

*All work was performed at the Argonne National Laboratory and supported by the Department of Energy, Office of Science, Materials Science and Engineering Division. The use of the Centre for Nanoscale Materials was supported by the US. Department of Energy (DOE), Office of Sciences, Basic Energy Sciences (BES), under Contract No. DE-AC02-06CH11357. Zhizhi Zhang acknowledges additional financial support from the China Scholarship Council (no. 201706160146) for a research stay at Argonne. José Holanda acknowledges additional financial support from the Conselho Nacional de Desenvolvimento Científico e Tecnológico(CNPq)-Brasil for a research stay at Argonne. Wei Zhang acknowledges support from U.S. National Science Foundation under Grants No. DMR-1808892.*

# Supplementary Movies

**Supplementary Movies1.** Corresponding animation of the first three odd modes ($n=1, 3, 5$) and the summation of the odd modes ($n=1, 3, 5, 7, 9, 11$) spatial normalized magnetization distribution in YIG microstripe as shown Fig. 1(c) in one period.

**Supplementary Movies2-5.** Corresponding animation of the first three modes ($n=1, 2, 3$) and the summation of all the first eleven modes spatial normalized magnetization distribution with $\Delta\varphi=0$ (**Movie2**), $\pi/2$ (**Movie3**), $\pi$ (**Movie4**), and $3\pi/2$ (**Movie5**) respectively in YIG microstripe as shown Fig. 1(e) in one period.

**Supplementary Movies6-12.** Corresponding animation of simulated spatial $m_z/M_s$ distribution in Fig. 2(c)-(i) with total time 80ns, respectively. The most brightness and darkness indicate the most positive and negative values respectively.

# Experiment

**Sample fabrication:** Micro-structured YIG stripe and the lateral Py dot were deposited on commercial polished (111)-oriented gadolinium gallium garnet ($Gd_3Ga_5O_{12}$, GGG). YIG was RF magnetron sputtered at room temperature (RT) from a stoichiometric YIG target. The Ar gas flow, chamber pressure, and sputtering power were kept at 16 sccm, 10 mTorr, and 75 W respectively. The microstructures were defined using electron beam lithography (Raith 150) on PMGI/ZED520 bilayer resists, which created an undercut cross-section profile. Since GGG is an insulator, a 5 nm Au layer was DC sputtered on the resists to avoid charge effects during electron beam exposure. Before the development, the Au was removed by exposure in gold etcher. And the electron beam exposed resists were developed in ZEDN50 (for ZED520) and 101A (for PMGI) developer respectively. After the deposition of YIG, the resist was lift-off by Shipley 1165 with only the microstripe structures left. The YIG was subsequently annealed ex situ at 850 °C for 3 h in a tube furnace, with ramped up time of 6 h and ramped down time of 14 h. After the YIG microstripe fabrication, the coplanar waveguide with shortened end made of Ti(20 nm)/Au(500 nm) was fabricated via optical lithography. After the μ-BLS measurement on the single YIG stripe, the Py dot was DC magnetron sputtered laterally near the YIG stripe, followed by the same electron beam lithography process. The precise alignment was performed in this step. The corresponding continuous YIG film and Py film capped with $SiO_2$ (15nm) on the whole substrates were also fabricated using the same process and fabrication parameters to characterize the material features.

**Material characterization:** The flip-chip vector network analyzer ferromagnetic resonance (VNA-FMR) method (Fig. S 1) was applied on the continuous films extended on the whole substrates to characterize the magnetic properties. We measured the transmission coefficient by sweeping the frequency at every fixed field. Therefore, the frequency swept linewidths ($\Delta f_{VNA}$) were obtained via Lorentz fitting. Detailed steps, including the conversion from $\Delta f_{VNA}$ to $\Delta H$, were referred to ref.[1] . The resonance

frequencies as a function of the magnetic field were fitted according to Kittel's equation:

$$f_{res} = \gamma\sqrt{H(H+M_s)} \qquad (1)$$

where the $M_s$ was yielded. And the $\alpha$ can be obtained through the fit:

$$\Delta H = \frac{2\alpha f_{res}}{\gamma} + \Delta H_0 \qquad (2)$$

where $\Delta H_0$ denotes the inhomogeneous linewidth broadening. Fig. S 2 depicts the magnetic properties of the magnetic films in the experiment.

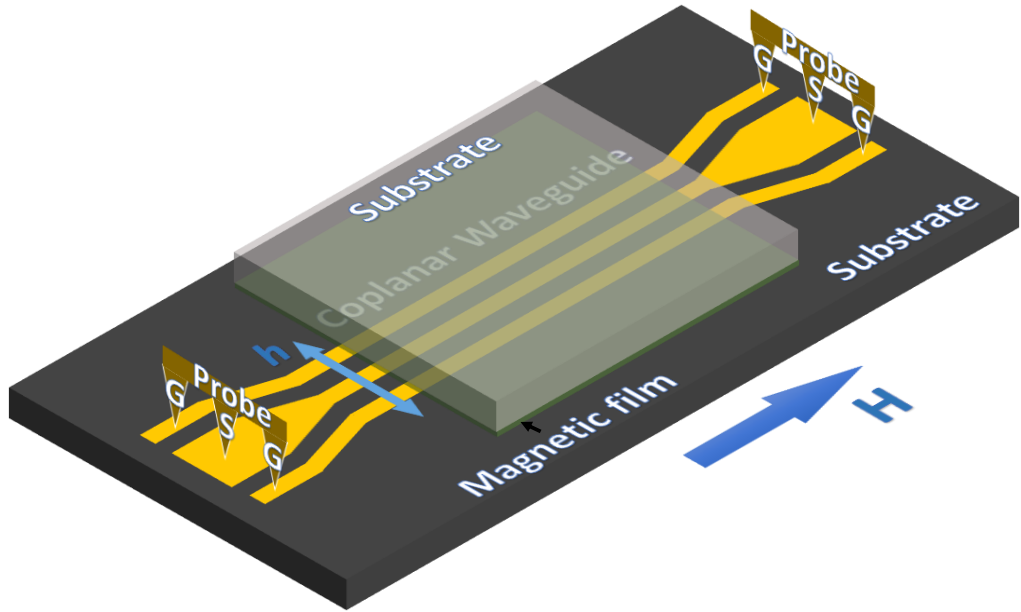

Fig. S 1 Schematic diagram of the VNA-FMR. The continuous magnetic films were placed on the coplanar waveguide structure. The applied external magnetic static field $H$ was perpendicular to the microwave field $h$.

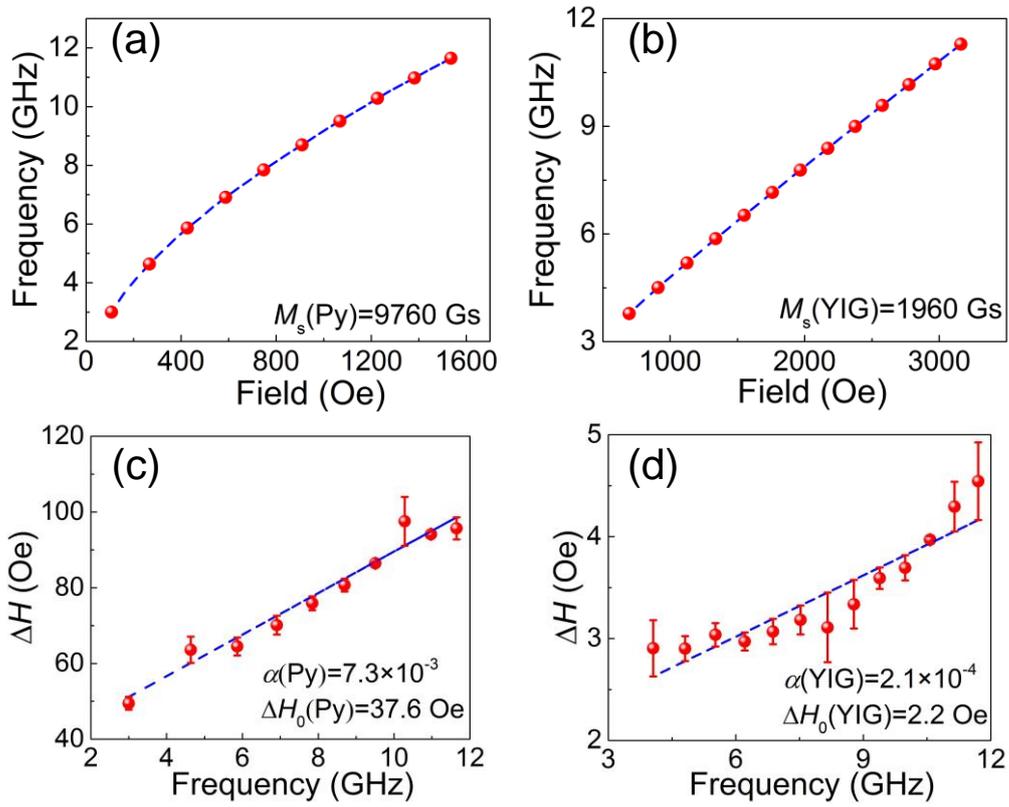

Fig. S 2 (a)Py and (b)YIG FMR frequency as a function of magnetic field. Error bars are smaller than the symbol size. (c)Py and (d)YIG FMR linewidth Δ$H$ as a function of the resonance frequency